\documentclass[reprint,
 amsmath,amssymb,
 aps,
pra,
]{revtex4-2}

\usepackage[english]{babel}
\usepackage{graphicx}
\usepackage{dcolumn}
\usepackage{bm}

\usepackage{amsmath, braket} 
\usepackage{amssymb}
\usepackage{bbold}
\usepackage{hyperref}
\usepackage{comment}
\usepackage{tikz}  
\usepackage[caption=false]{subfig}
\usetikzlibrary{arrows,shapes,positioning,shadows,svg.path,backgrounds,fit}

\begin{document}
\preprint{APS/123-QED}

\title{Non-Gaussianity and security of entanglement-based QKD}
\author{Mariia Gumberidze}
\email{gumberidze@optics.upol.cz}
\author{Vladyslav C. Usenko}
\email{usenko@optics.upol.cz}
\affiliation{
Department of Optics, Palack\'y University, 17. listopadu 12, 771 46 Olomouc, Czech Republic
}

\begin{abstract} 

We theoretically analyse the relation between non-Gaussianity and security of entanglement-based quantum key distribution (QKD) protocols, namely device-independent (DI) and entanglement-based BB$84$. A similar analysis has already been made for prepare-and-measure (P\&M) protocols \cite{Lasota2017}. In addition, we consider imperfect detection with dark counts and limited efficiency. We assume a perfect source of entangled Bell states as produced by quantum-dot type sources, depolarisation in the channel and different noise statistics, namely thermal and Poissonian. We consider single-photon avalanche photodiodes (SPAD) and photon number resolving detectors (PNRD) and use their respective criteria for non-Gaussianity. The results show cross-regions for both security and non-Gaussianity, hence, the possibility to conclude about the suitability of a given channel for secret key distribution. Our results can be useful as a pre-check for the implementation of QKD protocols.

\end{abstract}

\maketitle
\section{\label{intro} Introduction}

Discrete-variable (DV) quantum key distribution (QKD) protocols, which encode information in single photon or other quantum states with discrete spectrum, are particularly valued for their ability to maintain security over long distances despite channel losses and environmental noise \cite{Lasota2017}. Compared to continuous-variable (CV) protocols that rely on measuring field quadratures, DV-QKD schemes, especially their entanglement-based implementations, are more resilient against typical loss mechanisms encountered in fiber and free-space communication channels \cite{PhysRevA.76.012307}. A detailed comparison of the robustness of entanglement-based DV-QKD versus CV-QKD under thermal noise and loss conditions has been recently presented \cite{Lasota_2023}. Building upon this foundation, our work seeks to extend these studies by incorporating tests for the non-Gaussianity of light, previously applied to prepare-and-measure (P\&M) DV-QKD schemes \cite{Lasota2017}.

The non-Gaussian character of light serves as a significant indicator of its quantum nature. Several detection-based criteria have been formulated to identify the non-Gaussianity of optical states \cite{Filip2011, Lachman2013}. More recently, these criteria have been generalized to identify non-Gaussian coincidences in two-mode light fields \cite{Lachman2021, liu2023experimental}, which are instrumental in the analysis of bipartite quantum states, including entangled states.

In our work, we explore whether the non-Gaussianity of the signals in entanglement-based QKD can be a sufficient criterion for judging its applicability over given channel. Section~\ref{prereq} provides an overview of the prerequisites and relevant quantities considered in our analysis. In Section~\ref{noise}, we present the results of both non-Gaussianity and security analyses, accompanied by a visual comparison of the outcomes. The concluding remarks are presented in Section~\ref{conclusions}.

\begin{figure}
\begin{tikzpicture}
\node (img1){\includegraphics[width=\columnwidth]{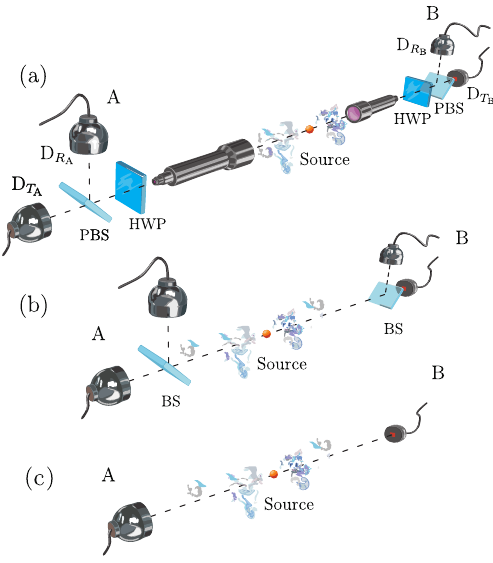}};
\node[above=of img1, node distance=0cm, yshift=-6.2cm,xshift=.5cm]
{\begin{minipage}{.3\textwidth}
 \begin{center}
 {\color{orange} SPAD}
 \end{center}
 \end{minipage}};
 \node[above=of img1, node distance=0cm, yshift=-9.8cm,xshift=2.5cm]
{\begin{minipage}{.3\textwidth}
 \begin{center}
 {\color{orange} PNRD}
 \end{center}
 \end{minipage}};
\end{tikzpicture}
\caption{\label{scheme} Schematics are shown for (a) the typical setup used in entanglement-based QKD protocols, consisting of polarization analyzers on each side comprising a half-wave plate (HWP), polarizing beam splitters (PBS), and two detectors $D_T, D_R$, where $T$ and $R$ stand for transmitted and reflected lights accordingly; (b) the setup for detecting non-Gaussian coincidences, which includes a balanced beam-splitter (BS) with transmittance $T=1/2$ and a pair of SPAD detectors on both Alice's and Bob's sides; (c) the setup for detecting non-Gaussian coincidences using PNRD on each side \cite{Lachman2021, liu2023experimental}. Different criteria, as given by \eqref{non-gauss-crit-spad} and \eqref{non-gauss-crit-pnrd}, are applied depending on the type of detectors.}
\end{figure}

\section{\label{prereq} Prerequisites}

\subsection{\label{key rates} 
Dewetak-Winter key rates for entanglement-based QKD} 

Dewetak-Winter key rates for DI-QKD and entanglement-based BB$84$ against collective attacks are given by \cite{Acin2007PRL, Pironio_2009}  
\begin{gather}
    r_{\text{dw}}^{\textit{DI-QKD}} \geq 1-h(Q)-h(\frac{1+\sqrt{(S/2)^{2}-1}}{2}),\\
    r_{\text{dw}} ^{\textit{BB}84} \geq 1-h(Q)-h(Q+\frac{S}{2\sqrt{2}}),
\end{gather}
where $h(Q)=\log_{2}(1-Q)-\log_{2}(Q)$ - binary entropy function, $Q$ is the QBER and $S$ is the Bell parameter.

We consider a special case of correlations $S=2\sqrt{2}(1-2Q)$. In this case, Alice measures in the basis $A_1=(\hat{\sigma}_{z}+\hat{\sigma}_{x})/2$, $A_2=(\hat{\sigma}_{z}-\hat{\sigma}_{x})/2$ and Bob's measurements are $B_1=\hat{\sigma}_{z}$, $B_2=\hat{\sigma}_{x}$, that maximise the Bell parameter $S$ for a noisy two-photon polarisation state $\hat{\rho}_{AB}= p \ket{\Phi^{+}}\bra{\Phi^{+}}+(1-p)/4\;\hat{\mathrm{I}}\otimes\hat{\mathrm{I}}$. The typical setup of entanglement-based QKD is shown in the Fig.~\ref{scheme}, (a).

\subsection{\label{photo-theory} Photodetection theory}

In practical setups, detection is imperfect due to limited efficiencies and the occurrence of dark counts. The following expressions describe the imperfect detection processes for SPAD and PNRD detectors.

\paragraph{SPAD.} In the case of SPAD detectors, the Positive Operator-Valued Measures (POVM) for a click and no-click of a respective detector are given by \cite{dusek2002koncepcni} 
\begin{gather}
\label{SPAD}
\hat{\Pi}_{c}=\displaystyle{\sum_{n=1}^{\infty}\left[1 - e^{-\nu}\left(1-\eta\right)^{n}\right]\ket{n}\bra{n}},\\
\hat{\Pi}_{0}=\displaystyle{e^{-\nu} \sum_{n=0}^{\infty}\left(1-\eta\right)^{n}\ket{n}\bra{n}},
\end{gather}
where $\eta$ denotes the efficiency of the detector and $\nu$ denotes the dark count rate. $\ket{n}$ denotes the $n$ Fock state.
\paragraph{PNRD.} In case of PNRD the POVM \cite{PNRD} for $n$-photon detection
\begin{equation}
\label{PNRD-n}
\hat{\Pi}_{n}=\displaystyle{e^{-\nu}\sum_{l=0}^{n} \sum_{k=n-l}^{\infty} \frac{\nu^l}{l!} C^{k}_{n-l}\eta^{n-l}(1-\eta)^{k-(n-l)}\ket{k}\bra{k}},
\end{equation}
where $C^{k}_{n-l}$ is the binomial coefficient,  $\{\eta, d\}$ denote the efficiency of the detector and dark counts rate, respectively.

In our analysis, the no-click detection 
\begin{equation}
\label{PNRD-0}
    \hat{\Pi}_{0}=\displaystyle{e^{-\nu} \sum_{k=0}^{\infty}  C^{k}_{0}(1-\eta)^{k}\ket{k}\bra{k}},
\end{equation}
and one-photon clicks are relevant
\begin{multline}
\label{PNRD-1}
\hat{\Pi}_{1}=\displaystyle{e^{-\nu}\sum_{l=0}^{1} \sum_{k=1-l}^{\infty} \frac{\nu^l}{l!} C^{k}_{1-l}\eta^{1-l}(1-\eta)^{k-(1-l)}\ket{k}\bra{k}}=\\
=\displaystyle{\eta \,e^{-\nu} \sum_{k=1}^{\infty} C^{k}_{1}(1-\eta)^{k-1}\ket{k}\bra{k}}+\\
+\nu \,e^{-\nu}\sum_{k=0}^{\infty} C^{k}_{0}(1-\eta)^{k}\ket{k}\bra{k},
\end{multline}
together with 2 or more photon detection needed for non-Gaussianity criterion 
\begin{multline}
\hat{\mathbf{1}}-\left(\hat{\Pi}_{0}+\hat{\Pi}_{1}\right)=\displaystyle{1-\eta \, e^{-\nu} \sum_{k=1}^{\infty} k\,(1-\eta)^{k-1}\ket{k}\bra{k}}-\\
-e^{-\nu}(1+\nu)\sum_{k=0}^{\infty} (1-\eta)^{k}\ket{k}\bra{k},
\end{multline}
where the binomial coefficients were simplified.

\subsection{\label{approximations} Non-Gaussianity criteria}

The setup for the detection of non-Gaussian coincidences differs for SPAD and PNR detectors, as shown in Fig.~\ref{scheme}, (b) and (c) respectively. The criteria for the former and the latter also differ, and they are presented below.

\paragraph{SPAD.} 
The non-Gaussianity criterion for SPAD detectors is given by
\begin{equation}
P_{s} > \frac{1}{2}\sqrt{\frac{P_{e}}{8+P_{e}}}\bigg(2+P_{e}+\sqrt{P_{e}(8+P_{e})}\bigg),
\label{non-gauss-crit-spad}
\end{equation}
where $P_{s}$ is the probability of a successful coincidence,, defined as the simultaneous click of a single detector on both Alice's and Bob's sides. $P_{e}$ is the probability of error coincidences, defined as $(P_{e}^{A}+P_{e}^{B})/2$, where $P_{e}^{A}$ and $P_{e}^{B}$ are independent events representing double clicks on Alice's or Bob's side, respectively.

\paragraph{PNRD.}
The analogous criterion for PNRD is expressed as
\begin{equation}
P_{s} > \sqrt{P_{e}}-P_{e},
\label{non-gauss-crit-pnrd}
\end{equation}
where $P_{s}$ is the probability of simultaneous detection of one photon on both Alice's and Bob's sides. $P_{e}$ is the average of error probabilities, $(P_{e}^{A}+P_{e}^{B})/2$, where $P_{e}^{A}$ and $P_{e}^{B}$ represent independent events corresponding to cases where two or more photons are registered on Alice's or Bob's side, respectively.

\begin{figure}
\begin{tikzpicture}
\node (img1){\includegraphics[width=0.46
\columnwidth]{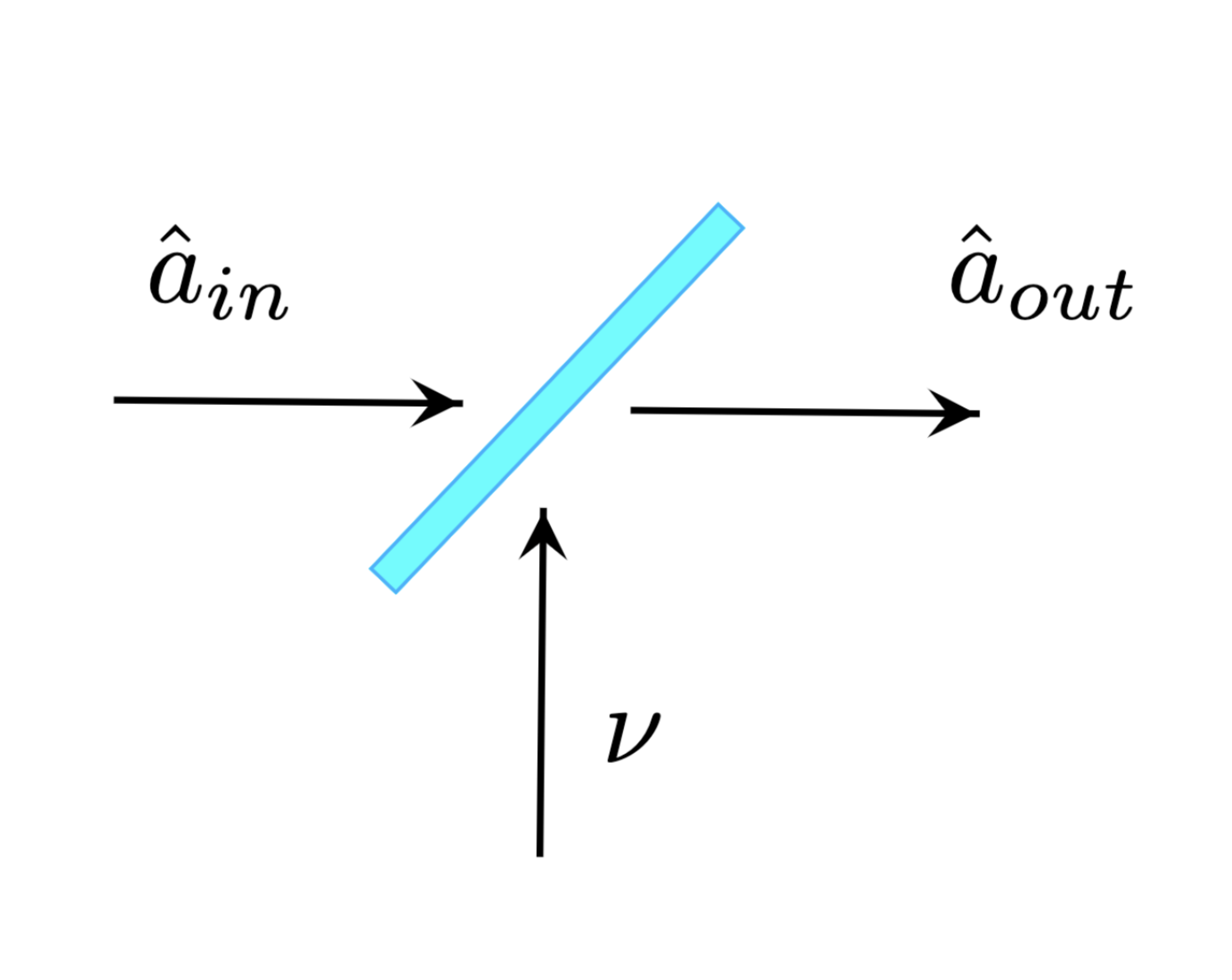}};
\node[right=of img1, node distance=0cm, yshift=-0.465cm,xshift=-1.5cm]
{\begin{minipage}{0.5\columnwidth}{\includegraphics[width=\columnwidth]{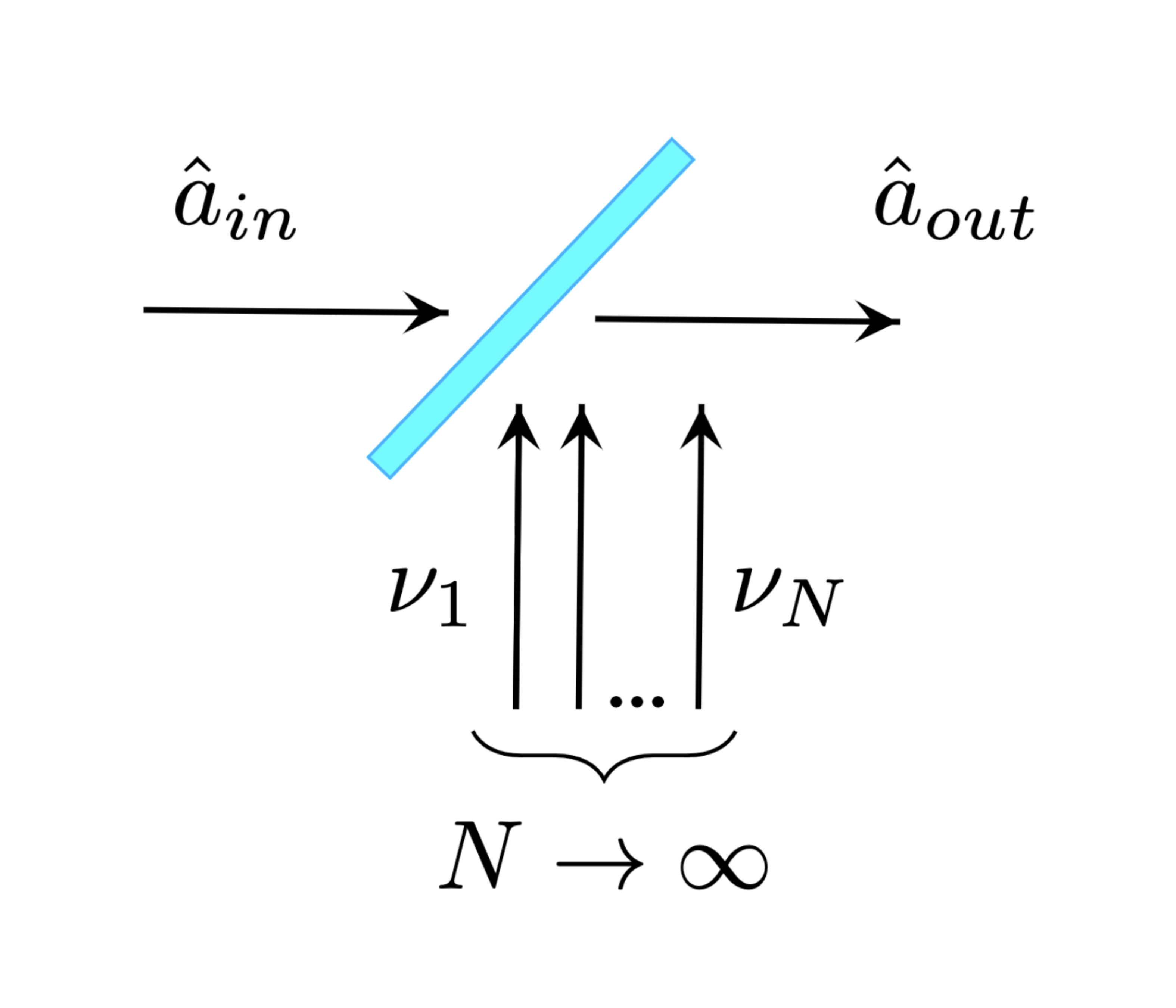}}
 \end{minipage}};
\node[above=of img1, node distance=0cm, yshift=-1.3cm,xshift=-0.3cm]{(a)};
\node[above=of img1, node distance=0cm, yshift=-1.3cm,xshift=3.9cm]{(b)};
\end{tikzpicture}
\caption{Model illustrating noise addition to the signal, incorporating two distinct noise types: (a) thermal noise characterized by thermal statistics with mean photon number $\nu$, and (b) multi-mode thermal noise which converges to Poissonian statistics in the limit of infinite number of modes. The effective coupling is modeled by the beamsplitter with transmittance $T$.}
\label{scheme-therm-poiss}
\end{figure}

\section{\label{noise} Noise addition and bounds}

The analysis is performed for two protocol types: entanglement-based BB$84$ and  DI-QKD. We assume that Alice and Bob can replace polarization analyzers with a non-Gaussianity detection setup at any time before or during QKD, as suggested in \cite{Lasota2017}. In addition, only collective attacks are considered. 

We assume the source deterministically prepares an entangled photon pair in the state $\ket{\Phi^{+}}=\frac{1}{\sqrt{2}}(\ket{00}+\ket{11})$. 
After passing through the depolarization channel, the state transforms into the Werner state
\begin{equation}
    \hat{\rho}(p)= p \ket{\Phi^{+}}\bra{\Phi^{+}}+\frac{1-p}{4}\;\hat{\mathrm{I}}\otimes\hat{\mathrm{I}},
    \label{wern}
\end{equation}
where $p$ is the probability of a pure Bell state without noise, $\hat{\mathrm{I}}$ denotes the $2$-dimensional identity matrix.

As the non-Gaussianity criteria are sensitive specifically to noise affecting the photon number distribution, we consider communication channels subject to both noise and losses. The noise is modeled with various statistical distributions, particularly thermal and Poissonian types. Furthermore, the analysis accounts for imperfect detection processes characterized by dark counts and detector efficiencies below unity. The criteria have been demonstrated theoretically and experimentally to be primarily sensitive to noise altering the photon number distribution, effectively detecting non-Gaussian features even when practical imperfections such as detector inefficiencies and dark counts are present \cite{Filip2011, Lachman2013, Lachman2021}.

In the modeling of noise addition, the polarization states \(\ket{0}\) and \(\ket{1}\) are first separated using a polarizing beam splitter (PBS). Noise is then individually introduced into each polarization mode by mixing with an auxiliary noise mode via a beam splitter (BS) with transmittance \(T\), as depicted in Fig.~\ref{scheme-therm-poiss}(a). For simplicity, this process is assumed to be symmetric and occurs equally on both sides, labeled A and B. After this noise addition process, the polarization modes are recombined to form the transmitted state. This general noise model, applicable to all QKD protocols tested for non-Gaussianity, captures the impact of various types of noise on the polarization components during transmission.

\subsection{\label{thermal} Thermal noise}

In this subsection, we examine the case where the signal gets coupled to noise modeled as a single thermal mode. This approach was originally used to describe imperfect detection \cite{Semenov2008}. 

\begin{figure}
\vspace{0.5cm}
\begin{tikzpicture} 
\node (img1)  {\includegraphics[width=.75\linewidth]{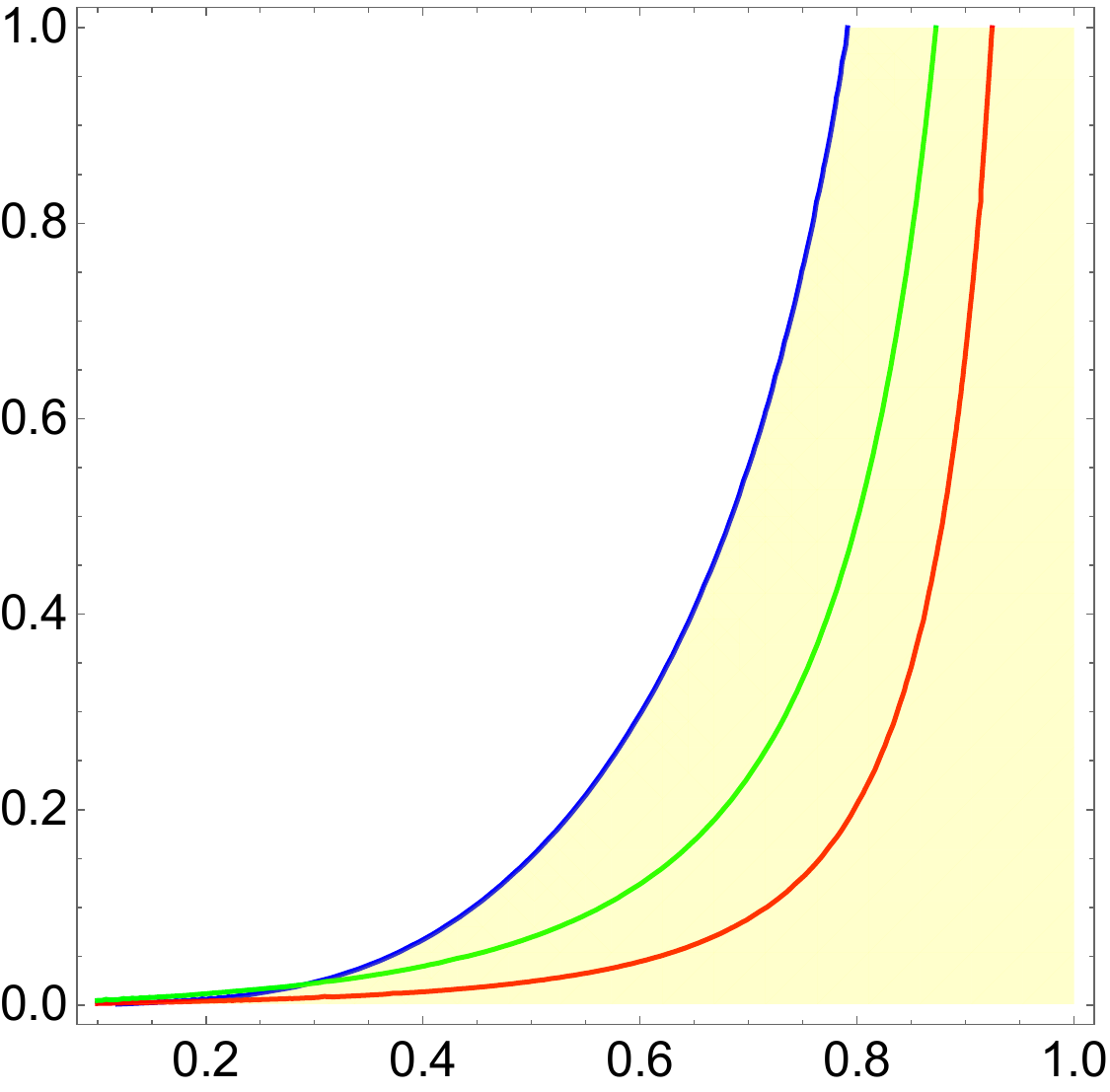}};
\node[above=of img1, node distance=0cm, rotate=90,yshift=3.4cm,xshift=-4.2cm] {$\nu$};
\node[above=of img1, node distance=0cm, yshift=-8.cm,xshift=0cm] {$T$};
\node[above=of img1, node distance=0cm, yshift=-1.9cm,xshift=0cm] {{\color{orange}$\textit{PNRD}$}};
\node[above=of img1, node distance=0cm, rotate=65,yshift=-3.2cm,xshift=-3.9cm] {{\color{blue}$\textit{non-Gauss}$}};
\node[above=of img1, node distance=0cm, rotate=55,yshift=-4.7cm,xshift=-4.cm] {{\color{green}$\textit{BB84}$}};
\node[above=of img1, node distance=0cm, rotate=50,yshift=-5.7cm,xshift=-3.7cm] {{\color{red}$\textit{DI-QKD}$}};
\end{tikzpicture}
\caption{\label{therm-perf} The maximum mean number of photons in a thermal mode $\nu$, compatible with non-Gaussianity and security of QKD, versus the coupling ratio $T$ between the signal and the noise modes upon perfect PNRD detection. The non-Gaussianity region is plotted by the yellow area bounded by the blue line. The Dewetak-Winter key rates for entanglement-based BB$84$ and DI-QKD protocols are plotted by the green and red lines, respectively. The cross-region of non-Gaussianity and security is clearly observed for coupling $T$ values larger than approx. $0.3$.}
\end{figure}
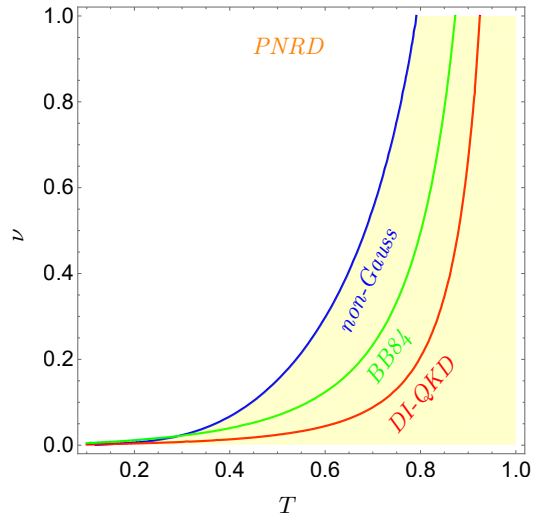


\paragraph{Perfect detection.} 
First, for simplicity, we consider the ideal case of perfect detection and thus focus on the photocount statistics of the signal after the beam splitter (BS), where it is mixed with a noise mode. The probability of detecting \(s\) counts \cite{Usenko2010} assuming the incident signal is in the Fock state \(\ket{l}\) is given by

\begin{multline}
     p_{sl}=Tr\left[\hat{U} \hat{\rho} \otimes \hat{\rho}_{th} \,\hat{U}^{\dagger} \ket{s}\bra{s} \otimes \mathbf{1} \right] =\\
    \sum_{n=0}^{\infty}\frac{1}{\bar{n}+1}\left(\frac{\bar{n}}{\bar{n}+1}\right)^{n} \left( \sum_{k=0}^{s} A^{l,n}_{k,s-k}\right)^2 \theta(l+n-s).
\end{multline}
Here, \(\rho = \ket{l}\bra{l}\) represents the signal state, \(\theta\) denotes the Heaviside step function, and the thermal state \(\hat{\rho}_{th}\) is given by

\[
\hat{\rho}_{th} = \frac{1}{\bar{n} + 1} \sum_{n=0}^\infty \left(\frac{\bar{n}}{\bar{n}+1}\right)^n \ket{n}\bra{n},
\]
where \(\bar{n}\) is the mean photon number. Since we consider a source generating a perfect Bell pair that only undergoes depolarization, we assume the signal state entering the beam splitter on both Alice’s and Bob’s sides is the Fock state \(\ket{1}\), thus effectively \(l=1\). Accordingly, the coefficient \(A\) is defined as

\begin{widetext}
\begin{equation}
      A^{l,n}_{k,s-k}=\sqrt{\frac{s!(l+n-s)!}{l!n!}}(-1)^{s-k}
     \begin{pmatrix}
     l\\ k
     \end{pmatrix}
     \begin{pmatrix}
     n\\ s-k
     \end{pmatrix}
     (\sqrt{1-T})^{l+s-2k}(\sqrt{T})^{n+2k-s},
\end{equation}
\end{widetext}
where $T$ stands for the transmittance of the BS.

In this case, we restrict the analysis to PNRDs, which allows the direct application of the photocounting statistics \cite{Usenko2010} described above. The security assessment of entanglement-based QKD protocols relies on an accurate evaluation of QBER,, as the relation between the Bell parameter and QBER, \( S = 2 \sqrt{2} (1 - 2Q) \), is retained. The QBER in the corresponding notation reads
\begin{equation}
    Q = \frac{4\, p \, p_{11} \, p_{00} \, p_{01} \, p_{10} + (1 - p) N}{2N},
\end{equation}
where
\begin{equation}
    N = (p_{11} \, p_{00} + p_{10} \, p_{01})^2
\end{equation}
represents the white noise term in the Werner state \eqref{wern}. Similarly, the probabilities relevant for the non-Gaussianity criterion with PNRD detectors are given by
\begin{equation}
    P_s = (p_{11})^2, \quad P_e = 1 - p_{01} - p_{11}.
\end{equation}

The results shown in Fig.~\ref{therm-perf} demonstrate that the non-Gaussianity pre-check can serve as a reliable indicator of protocol security in this scenario, particularly in the regime where the coupling ratio $T$ exceeds approximately $0.3$. Beyond this threshold, the correspondence between non-Gaussianity and security becomes increasingly pronounced, supporting the use of the non-Gaussianity criterion as an effective preliminary test for protocol security.

\paragraph{Imperfect detection.} 
In this section, we incorporate imperfect detection by applying the photodetection equations, introduced in Subsec.~\ref{photo-theory}. Although the expression for the QBER remains unchanged, the probabilities \(p_{sl}\) are modified accordingly to reflect the imperfections as
\begin{gather}
    \Tilde{p}_{0l}=e^{-\nu} \sum_{s=0}^{\infty}  C^{s}_{0}(1-\eta)^{s}\,p_{sl},\\
    \Tilde{p}_{1l}=\eta \,e^{-\nu} \sum_{s=1}^{\infty} C^{s}_{1}(1-\eta)^{s-1}\,p_{sl}+\nu \,e^{-\nu}\sum_{s=0}^{\infty} C^{s}_{0}(1-\eta)^{s}\,p_{sl},
\end{gather}
where Eq.~\eqref{PNRD-1},\eqref{PNRD-0} for imperfect PNRD were employed. Analogously, the non-Gaussianity probabilities \(P_s\) and \(P_e\) retain the same form, but are now expressed in terms of the modified probabilities \(\tilde{p}_{01}\) and \(\tilde{p}_{11}\).

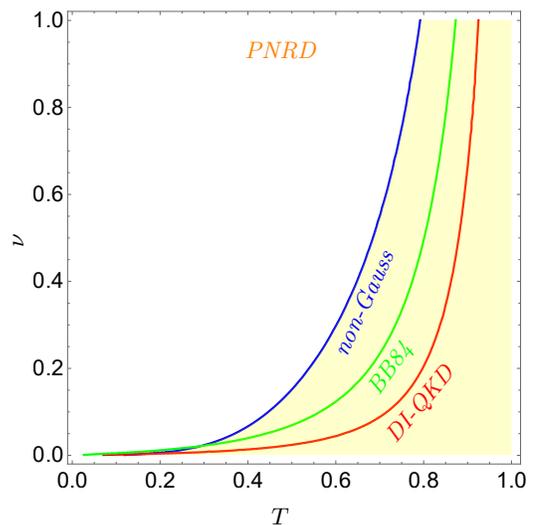
\begin{figure}
\vspace{0.5cm}
\begin{tikzpicture} 
\node (img1)  {\includegraphics[width=.76\columnwidth]{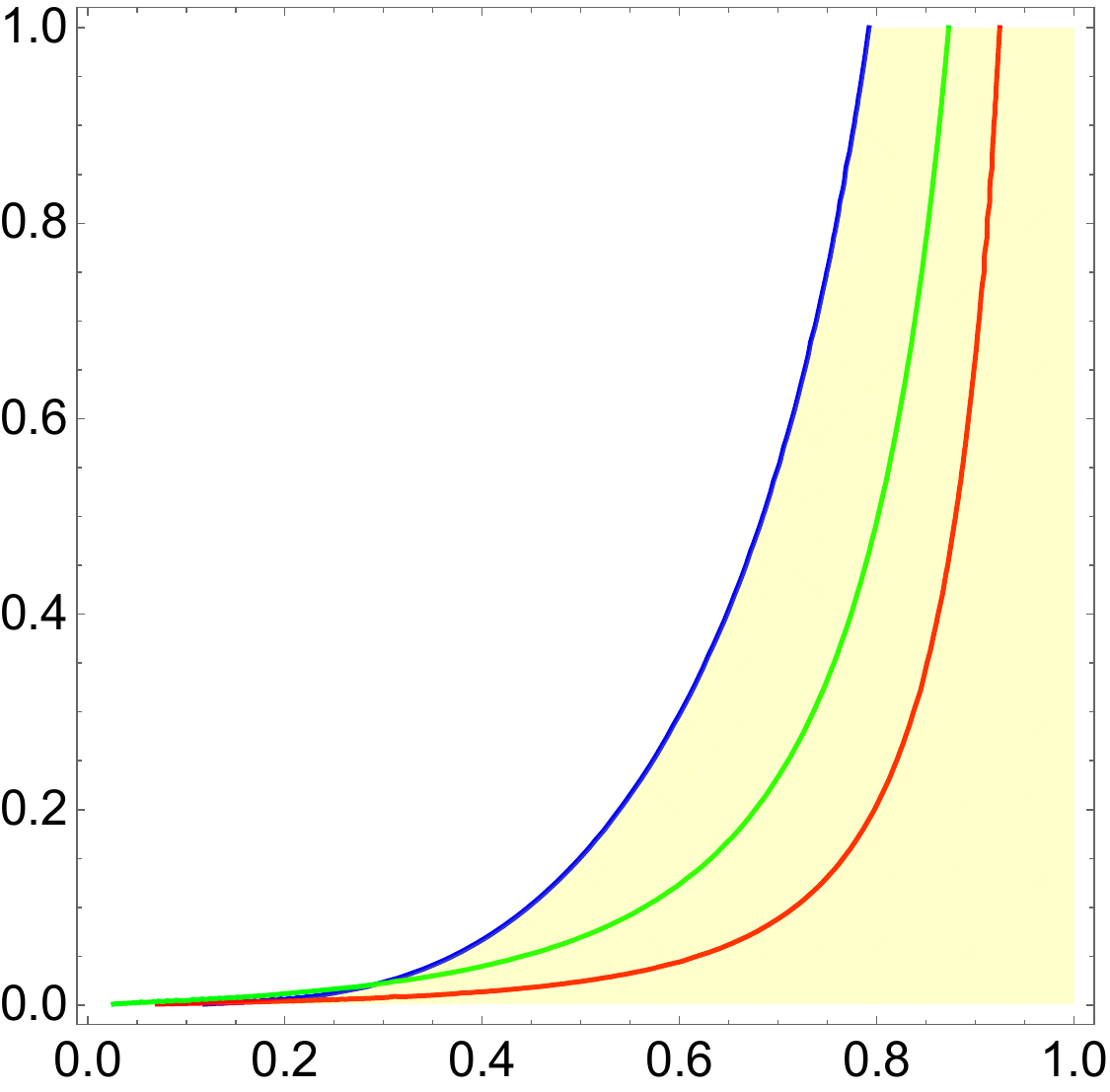}};
\node[above=of img1, node distance=0cm, rotate=90,yshift=3.3cm,xshift=-4.2cm] {$\nu$};
\node[above=of img1, node distance=0cm, yshift=-8.1cm,xshift=0cm] {$T$};
\node[above=of img1, node distance=0cm, yshift=-1.9cm,xshift=0cm] {{\color{orange}$\textit{PNRD}$}};
\node[above=of img1, node distance=0cm, rotate=65,yshift=-3.35cm,xshift=-4.1cm] {{\color{blue}$\textit{non-Gauss}$}};
\node[above=of img1, node distance=0cm, rotate=55,yshift=-4.85cm,xshift=-4.cm] {{\color{green}$\textit{BB84}$}};
\node[above=of img1, node distance=0cm, rotate=50,yshift=-5.8cm,xshift=-3.7cm] {{\color{red}$\textit{DI-QKD}$}};
\end{tikzpicture}
\caption{\label{therm-imperf} 
The maximum mean number of photons in a thermal mode $\nu$, compatible with non-Gaussianity and security of QKD, versus the coupling ratio \(T\) between the signal and noise modes upon imperfect PNRD detection with efficiency \(\eta = 0.7\) and dark count rate \(d = 0.001\). The non-Gaussianity region is highlighted by the yellow area bounded by the blue curve. Dewetak-Winter key rates for entanglement-based BB84 and DI-QKD protocols are shown as green and red lines, respectively. The regions corresponding to non-Gaussianity and security nearly overlap, except at low values of coupling \(T \lesssim 0.3\).}
\end{figure}

The results are presented in Fig.~\ref{therm-imperf} and exhibit the same qualitative behavior as those in Fig.~\ref{therm-perf}, indicating that the presence of imperfect detection does not significantly impede the estimation of non-Gaussianity. Notably, this consistency holds in the region where the coupling exceeds approximately $0.3$, reinforcing the robustness of the non-Gaussianity pre-check as a reliable indicator of protocol security under realistic detection conditions.

\subsection{\label{poissonian} Poissonian noise}

The interaction of the signal with multi-mode thermal light at a beam splitter with transmittance $T$, as illustrated in Fig.~\ref{scheme-therm-poiss}(b), is mathematically equivalent to imperfect detection with efficiency $\eta$ and dark count rate $d$ in the limit of an infinite number of thermal modes~\cite{Semenov2008}. Consequently, for analytical convenience, the effects of Poissonian channel noise can be incorporated into the detection model, i.e., the POVM operators. In this analysis, we focus on SPAD detectors \eqref{SPAD} to ensure the applicability of the chosen non-Gaussianity criteria.

\begin{figure}
\vspace{0.5cm}
\begin{tikzpicture} 
\node (img1)  {\includegraphics[width=.75\linewidth]{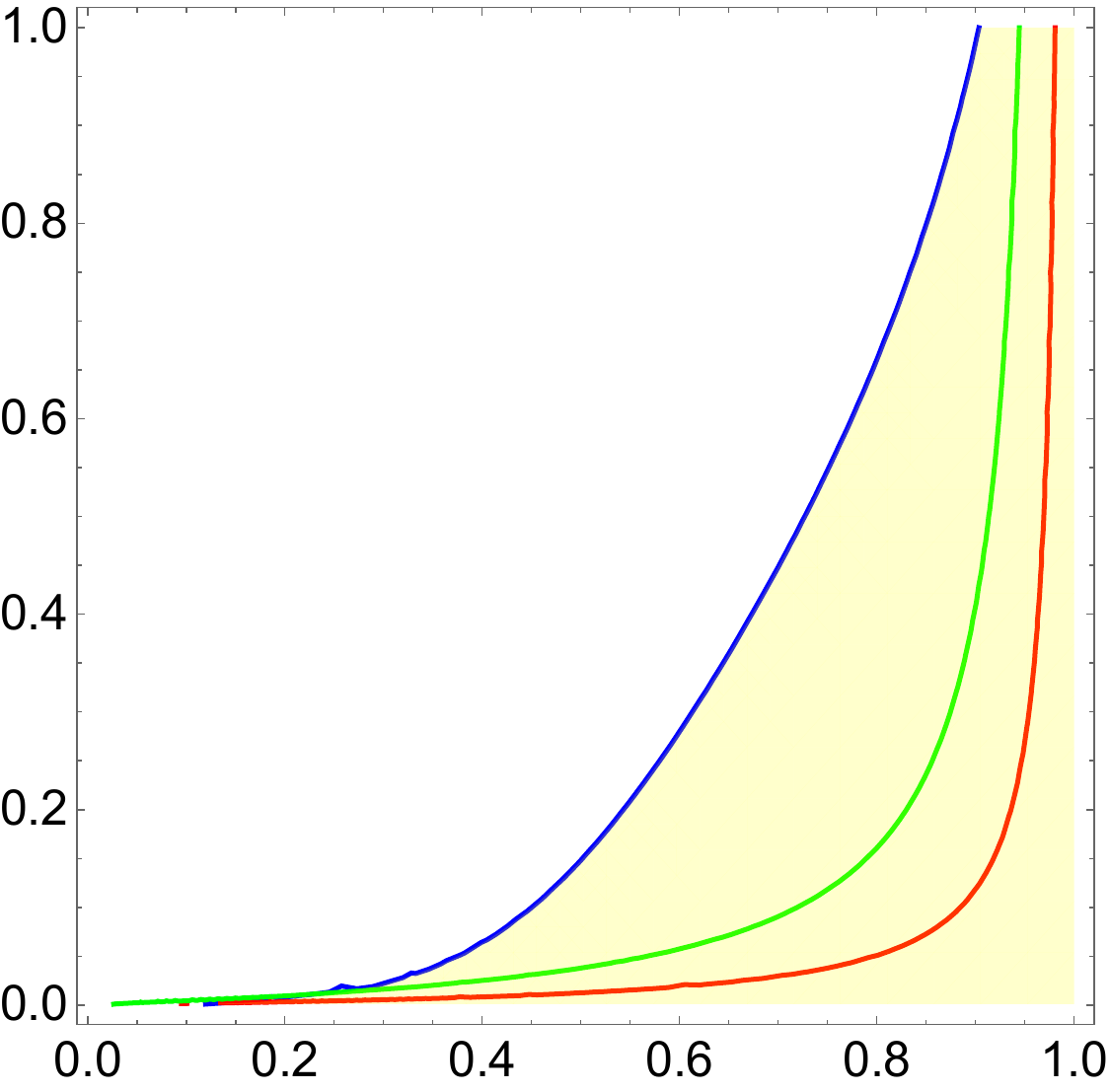}};
\node[above=of img1, node distance=0cm, rotate=90,yshift=3.3cm,xshift=-4.2cm] {$\nu$};
\node[above=of img1, node distance=0cm, yshift=-8.cm,xshift=0cm] {$T$};
\node[above=of img1, node distance=0cm, yshift=-1.9cm,xshift=0cm] {{\color{orange}$\textit{SPAD}$}};
\node[above=of img1, node distance=0cm, rotate=60,yshift=-3.8cm,xshift=-3.5cm] {{\color{blue}$\textit{non-Gauss}$}};
\node[above=of img1, node distance=0cm, rotate=55,yshift=-5.cm,xshift=-3.8cm] {{\color{green}$\textit{BB84}$}};
\node[above=of img1, node distance=0cm, rotate=50,yshift=-5.85cm,xshift=-3.3cm] {{\color{red}$\textit{DI-QKD}$}};
\end{tikzpicture}
\caption{\label{poisson-imperf} The maximum mean number of photons in a thermal mode $\nu$, compatible with non-Gaussianity and security of QKD, versus the coupling ratio $T$ upon imperfect SPAD detection with efficiency \(\eta\) and dark count rate \(d\). The results illustrate nearly complete overlap between the regions of security and non-Gaussianity of the detected light, confirming the robustness of the non-Gaussianity criterion under realistic detection conditions. Inconclusive behavior is observed in the region of low coupling \(T \lesssim 0.3\).}
\end{figure}
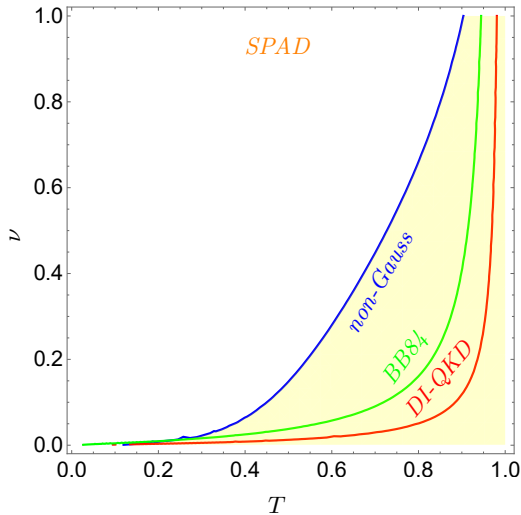

QBER quantifies the rate of anti-correlations between the measurement outcomes of the two communicating parties and results in
\begin{equation}
    Q = \frac{1}{2} - \frac{e^{2\,d}\,p\, \eta^2}{2\,(2 + e^d\,(\eta -2) - 2\, \eta)^2},
\end{equation}
whereas success and error probabilities associated with the non-Gaussianity criterion \eqref{non-gauss-crit-spad} amount to 
\begin{gather}
    P_{s} = \frac{1}{4} 
\,e^{-4\,d}\,(2 + e^n (\eta -2) - 2 \eta)^2,\\
    P_{e} = e^{-4\,d}\,(1 - e^d)\,(1-\eta)\,(1-\eta - e^d).
\end{gather}
Remarkably, since four distinct coincidence outcomes qualify as successful events for the non-Gaussianity test, we compute their total contribution and report the average by dividing by $4$.

The corresponding results are shown in Fig.~\ref{poisson-imperf}. Although the non-Gaussianity region differs slightly from the thermal-noise case with PNRD, the findings still support that estimating non-Gaussianity is a reliable indicator of potential QKD security.

\section{\label{conclusions} Conclusions and outlook} We investigated the relationship between quantum non-Gaussianity and security for entanglement-based QKD over depolarizing channels, considering two representative noise models (single-mode thermal and multimode Poissonian) and two detector types (SPAD and PNRD) with their respective non-Gaussianity witnesses. Across both DI-QKD and entanglement-based BB84, the analyses consistently reveal nonzero cross-regions in which the non-Gaussianity witness is satisfied and the Devetak–Winter key rate is positive, indicating that non-Gaussianity functions as a practical pre-check for channel suitability within an operational transmittance window. At high noise loads, the overlap between witness satisfaction and positive key rates diminishes, rendering the diagnostic inconclusive, which sets a natural operational boundary for the pre-check, specifically in the cases of low coupling ratio \(T \lesssim 0.3\).

Realistic detection imperfections—efficiency below unity and nonzero dark counts—were incorporated via POVMs, and results were reported for a standard set of CHSH-optimal correlations used in entanglement-based BB84 and DI-QKD. The observed qualitative behavior is robust to these imperfections: the cross-region between non-Gaussianity and security largely persists, with deviations mainly at low transmittance. While the figures here focus on a specific correlation choice, the methodology extends straightforwardly to other correlation subsets and measurement settings, enabling adaptation to diverse experimental configurations.

\begin{acknowledgments}

Authors acknowledge support from the project no. 21-44815L of the Czech Science Foundation, the project 8C22003 (QD-E-QKD) of MEYS of Czech Republic, which has received funding from the European Union’s Horizon 2020 research and innovation framework programme under Grant Agreement No. 731473 and 101017733, and ”Quantum Secure Networks Partnership” (QSNP, grant agreement No. 101114043) project funded from the European Union’s Horizon Europe research and innovation programme.

\end{acknowledgments}

\bibliography{apssamp}

\begin{thebibliography}{14}%
\makeatletter
\providecommand \@ifxundefined [1]{%
 \@ifx{#1\undefined}
}%
\providecommand \@ifnum [1]{%
 \ifnum #1\expandafter \@firstoftwo
 \else \expandafter \@secondoftwo
 \fi
}%
\providecommand \@ifx [1]{%
 \ifx #1\expandafter \@firstoftwo
 \else \expandafter \@secondoftwo
 \fi
}%
\providecommand \natexlab [1]{#1}%
\providecommand \enquote  [1]{``#1''}%
\providecommand \bibnamefont  [1]{#1}%
\providecommand \bibfnamefont [1]{#1}%
\providecommand \citenamefont [1]{#1}%
\providecommand \href@noop [0]{\@secondoftwo}%
\providecommand \href [0]{\begingroup \@sanitize@url \@href}%
\providecommand \@href[1]{\@@startlink{#1}\@@href}%
\providecommand \@@href[1]{\endgroup#1\@@endlink}%
\providecommand \@sanitize@url [0]{\catcode `\\12\catcode `\$12\catcode `\&12\catcode `\#12\catcode `\^12\catcode `\_12\catcode `\%12\relax}%
\providecommand \@@startlink[1]{}%
\providecommand \@@endlink[0]{}%
\providecommand \url  [0]{\begingroup\@sanitize@url \@url }%
\providecommand \@url [1]{\endgroup\@href {#1}{\urlprefix }}%
\providecommand \urlprefix  [0]{URL }%
\providecommand \Eprint [0]{\href }%
\providecommand \doibase [0]{https://doi.org/}%
\providecommand \selectlanguage [0]{\@gobble}%
\providecommand \bibinfo  [0]{\@secondoftwo}%
\providecommand \bibfield  [0]{\@secondoftwo}%
\providecommand \translation [1]{[#1]}%
\providecommand \BibitemOpen [0]{}%
\providecommand \bibitemStop [0]{}%
\providecommand \bibitemNoStop [0]{.\EOS\space}%
\providecommand \EOS [0]{\spacefactor3000\relax}%
\providecommand \BibitemShut  [1]{\csname bibitem#1\endcsname}%
\let\auto@bib@innerbib\@empty
\bibitem [{\citenamefont {Lasota}\ \emph {et~al.}(2017)\citenamefont {Lasota}, \citenamefont {Filip},\ and\ \citenamefont {Usenko}}]{Lasota2017}%
  \BibitemOpen
  \bibfield  {author} {\bibinfo {author} {\bibfnamefont {M.}~\bibnamefont {Lasota}}, \bibinfo {author} {\bibfnamefont {R.}~\bibnamefont {Filip}},\ and\ \bibinfo {author} {\bibfnamefont {V.~C.}\ \bibnamefont {Usenko}},\ }\bibfield  {title} {\bibinfo {title} {Sufficiency of quantum non-gaussianity for discrete-variable quantum key distribution over noisy channels},\ }\href {https://doi.org/10.1103/PhysRevA.96.012301} {\bibfield  {journal} {\bibinfo  {journal} {Phys. Rev. A}\ }\textbf {\bibinfo {volume} {96}},\ \bibinfo {pages} {012301} (\bibinfo {year} {2017})}\BibitemShut {NoStop}%
\bibitem [{\citenamefont {Ma}\ \emph {et~al.}(2007)\citenamefont {Ma}, \citenamefont {Fung},\ and\ \citenamefont {Lo}}]{PhysRevA.76.012307}%
  \BibitemOpen
  \bibfield  {author} {\bibinfo {author} {\bibfnamefont {X.}~\bibnamefont {Ma}}, \bibinfo {author} {\bibfnamefont {C.-H.~F.}\ \bibnamefont {Fung}},\ and\ \bibinfo {author} {\bibfnamefont {H.-K.}\ \bibnamefont {Lo}},\ }\bibfield  {title} {\bibinfo {title} {Quantum key distribution with entangled photon sources},\ }\href {https://doi.org/10.1103/PhysRevA.76.012307} {\bibfield  {journal} {\bibinfo  {journal} {Phys. Rev. A}\ }\textbf {\bibinfo {volume} {76}},\ \bibinfo {pages} {012307} (\bibinfo {year} {2007})}\BibitemShut {NoStop}%
\bibitem [{\citenamefont {Lasota}\ \emph {et~al.}(2023)\citenamefont {Lasota}, \citenamefont {Kovalenko},\ and\ \citenamefont {Usenko}}]{Lasota_2023}%
  \BibitemOpen
  \bibfield  {author} {\bibinfo {author} {\bibfnamefont {M.}~\bibnamefont {Lasota}}, \bibinfo {author} {\bibfnamefont {O.}~\bibnamefont {Kovalenko}},\ and\ \bibinfo {author} {\bibfnamefont {V.~C.}\ \bibnamefont {Usenko}},\ }\bibfield  {title} {\bibinfo {title} {Robustness of entanglement-based discrete- and continuous-variable quantum key distribution against channel noise},\ }\href {https://doi.org/10.1088/1367-2630/ad0e8c} {\bibfield  {journal} {\bibinfo  {journal} {New Journal of Physics}\ }\textbf {\bibinfo {volume} {25}},\ \bibinfo {pages} {123003} (\bibinfo {year} {2023})}\BibitemShut {NoStop}%
\bibitem [{\citenamefont {Filip}\ and\ \citenamefont {Mi\ifmmode~\check{s}\else \v{s}\fi{}ta}(2011)}]{Filip2011}%
  \BibitemOpen
  \bibfield  {author} {\bibinfo {author} {\bibfnamefont {R.}~\bibnamefont {Filip}}\ and\ \bibinfo {author} {\bibfnamefont {L.}~\bibnamefont {Mi\ifmmode~\check{s}\else \v{s}\fi{}ta}},\ }\bibfield  {title} {\bibinfo {title} {Detecting quantum states with a positive wigner function beyond mixtures of gaussian states},\ }\href {https://doi.org/10.1103/PhysRevLett.106.200401} {\bibfield  {journal} {\bibinfo  {journal} {Phys. Rev. Lett.}\ }\textbf {\bibinfo {volume} {106}},\ \bibinfo {pages} {200401} (\bibinfo {year} {2011})}\BibitemShut {NoStop}%
\bibitem [{\citenamefont {Lachman}\ and\ \citenamefont {Filip}(2013)}]{Lachman2013}%
  \BibitemOpen
  \bibfield  {author} {\bibinfo {author} {\bibfnamefont {L.~c.~v.}\ \bibnamefont {Lachman}}\ and\ \bibinfo {author} {\bibfnamefont {R.}~\bibnamefont {Filip}},\ }\bibfield  {title} {\bibinfo {title} {Robustness of quantum nonclassicality and non-gaussianity of single-photon states in attenuating channels},\ }\href {https://doi.org/10.1103/PhysRevA.88.063841} {\bibfield  {journal} {\bibinfo  {journal} {Phys. Rev. A}\ }\textbf {\bibinfo {volume} {88}},\ \bibinfo {pages} {063841} (\bibinfo {year} {2013})}\BibitemShut {NoStop}%
\bibitem [{\citenamefont {Lachman}\ and\ \citenamefont {Filip}(2021)}]{Lachman2021}%
  \BibitemOpen
  \bibfield  {author} {\bibinfo {author} {\bibfnamefont {L.}~\bibnamefont {Lachman}}\ and\ \bibinfo {author} {\bibfnamefont {R.}~\bibnamefont {Filip}},\ }\bibfield  {title} {\bibinfo {title} {Quantum non-gaussian photon coincidences},\ }\href {https://doi.org/10.1103/PhysRevLett.126.213604} {\bibfield  {journal} {\bibinfo  {journal} {Phys. Rev. Lett.}\ }\textbf {\bibinfo {volume} {126}},\ \bibinfo {pages} {213604} (\bibinfo {year} {2021})}\BibitemShut {NoStop}%
\bibitem [{\citenamefont {Liu}\ \emph {et~al.}(2023)\citenamefont {Liu}, \citenamefont {Qiao}, \citenamefont {Lachman}, \citenamefont {Ge}, \citenamefont {Chung}, \citenamefont {Zhao}, \citenamefont {Li}, \citenamefont {You}, \citenamefont {Filip},\ and\ \citenamefont {Huo}}]{liu2023experimental}%
  \BibitemOpen
  \bibfield  {author} {\bibinfo {author} {\bibfnamefont {R.-Z.}\ \bibnamefont {Liu}}, \bibinfo {author} {\bibfnamefont {Y.-K.}\ \bibnamefont {Qiao}}, \bibinfo {author} {\bibfnamefont {L.}~\bibnamefont {Lachman}}, \bibinfo {author} {\bibfnamefont {Z.-X.}\ \bibnamefont {Ge}}, \bibinfo {author} {\bibfnamefont {T.-H.}\ \bibnamefont {Chung}}, \bibinfo {author} {\bibfnamefont {J.-Y.}\ \bibnamefont {Zhao}}, \bibinfo {author} {\bibfnamefont {H.}~\bibnamefont {Li}}, \bibinfo {author} {\bibfnamefont {L.}~\bibnamefont {You}}, \bibinfo {author} {\bibfnamefont {R.}~\bibnamefont {Filip}},\ and\ \bibinfo {author} {\bibfnamefont {Y.-H.}\ \bibnamefont {Huo}},\ }\href@noop {} {\bibinfo {title} {Experimental quantum non-gaussian coincidences of entangled photons}} (\bibinfo {year} {2023}),\ \Eprint {https://arxiv.org/abs/2307.04531} {arXiv:2307.04531 [quant-ph]} \BibitemShut {NoStop}%
\bibitem [{\citenamefont {Ac\'{\i}n}\ \emph {et~al.}(2007)\citenamefont {Ac\'{\i}n}, \citenamefont {Brunner}, \citenamefont {Gisin}, \citenamefont {Massar}, \citenamefont {Pironio},\ and\ \citenamefont {Scarani}}]{Acin2007PRL}%
  \BibitemOpen
  \bibfield  {author} {\bibinfo {author} {\bibfnamefont {A.}~\bibnamefont {Ac\'{\i}n}}, \bibinfo {author} {\bibfnamefont {N.}~\bibnamefont {Brunner}}, \bibinfo {author} {\bibfnamefont {N.}~\bibnamefont {Gisin}}, \bibinfo {author} {\bibfnamefont {S.}~\bibnamefont {Massar}}, \bibinfo {author} {\bibfnamefont {S.}~\bibnamefont {Pironio}},\ and\ \bibinfo {author} {\bibfnamefont {V.}~\bibnamefont {Scarani}},\ }\bibfield  {title} {\bibinfo {title} {Device-independent security of quantum cryptography against collective attacks},\ }\href {https://doi.org/10.1103/PhysRevLett.98.230501} {\bibfield  {journal} {\bibinfo  {journal} {Phys. Rev. Lett.}\ }\textbf {\bibinfo {volume} {98}},\ \bibinfo {pages} {230501} (\bibinfo {year} {2007})}\BibitemShut {NoStop}%
\bibitem [{\citenamefont {Pironio}\ \emph {et~al.}(2009)\citenamefont {Pironio}, \citenamefont {Acín}, \citenamefont {Brunner}, \citenamefont {Gisin}, \citenamefont {Massar},\ and\ \citenamefont {Scarani}}]{Pironio_2009}%
  \BibitemOpen
  \bibfield  {author} {\bibinfo {author} {\bibfnamefont {S.}~\bibnamefont {Pironio}}, \bibinfo {author} {\bibfnamefont {A.}~\bibnamefont {Acín}}, \bibinfo {author} {\bibfnamefont {N.}~\bibnamefont {Brunner}}, \bibinfo {author} {\bibfnamefont {N.}~\bibnamefont {Gisin}}, \bibinfo {author} {\bibfnamefont {S.}~\bibnamefont {Massar}},\ and\ \bibinfo {author} {\bibfnamefont {V.}~\bibnamefont {Scarani}},\ }\bibfield  {title} {\bibinfo {title} {Device-independent quantum key distribution secure against collective attacks},\ }\href {https://doi.org/10.1088/1367-2630/11/4/045021} {\bibfield  {journal} {\bibinfo  {journal} {New Journal of Physics}\ }\textbf {\bibinfo {volume} {11}},\ \bibinfo {pages} {045021} (\bibinfo {year} {2009})}\BibitemShut {NoStop}%
\bibitem [{\citenamefont {Dušek}(2002)}]{dusek2002koncepcni}%
  \BibitemOpen
  \bibfield  {author} {\bibinfo {author} {\bibfnamefont {M.}~\bibnamefont {Dušek}},\ }\href@noop {} {\emph {\bibinfo {title} {Koncepční otázky kvantové teorie}}}\ (\bibinfo  {publisher} {Univerzita Palackého},\ \bibinfo {address} {Olomouc},\ \bibinfo {year} {2002})\ p.\ \bibinfo {pages} {236}\BibitemShut {NoStop}%
\bibitem [{\citenamefont {Izumi}\ \emph {et~al.}(2013)\citenamefont {Izumi}, \citenamefont {Takeoka}, \citenamefont {Ema},\ and\ \citenamefont {Sasaki}}]{PNRD}%
  \BibitemOpen
  \bibfield  {author} {\bibinfo {author} {\bibfnamefont {S.}~\bibnamefont {Izumi}}, \bibinfo {author} {\bibfnamefont {M.}~\bibnamefont {Takeoka}}, \bibinfo {author} {\bibfnamefont {K.}~\bibnamefont {Ema}},\ and\ \bibinfo {author} {\bibfnamefont {M.}~\bibnamefont {Sasaki}},\ }\bibfield  {title} {\bibinfo {title} {Quantum receivers with squeezing and photon-number-resolving detectors for $m$-ary coherent state discrimination},\ }\href {https://doi.org/10.1103/PhysRevA.87.042328} {\bibfield  {journal} {\bibinfo  {journal} {Phys. Rev. A}\ }\textbf {\bibinfo {volume} {87}},\ \bibinfo {pages} {042328} (\bibinfo {year} {2013})}\BibitemShut {NoStop}%
\bibitem [{\citenamefont {Semenov}\ \emph {et~al.}(2008)\citenamefont {Semenov}, \citenamefont {Turchin},\ and\ \citenamefont {Gomonay}}]{Semenov2008}%
  \BibitemOpen
  \bibfield  {author} {\bibinfo {author} {\bibfnamefont {A.~A.}\ \bibnamefont {Semenov}}, \bibinfo {author} {\bibfnamefont {A.~V.}\ \bibnamefont {Turchin}},\ and\ \bibinfo {author} {\bibfnamefont {H.~V.}\ \bibnamefont {Gomonay}},\ }\bibfield  {title} {\bibinfo {title} {Detection of quantum light in the presence of noise},\ }\href {https://doi.org/10.1103/PhysRevA.78.055803} {\bibfield  {journal} {\bibinfo  {journal} {Phys. Rev. A}\ }\textbf {\bibinfo {volume} {78}},\ \bibinfo {pages} {055803} (\bibinfo {year} {2008})}\BibitemShut {NoStop}%
\bibitem [{\citenamefont {Usenko}\ and\ \citenamefont {Paris}(2010)}]{Usenko2010}%
  \BibitemOpen
  \bibfield  {author} {\bibinfo {author} {\bibfnamefont {V.~C.}\ \bibnamefont {Usenko}}\ and\ \bibinfo {author} {\bibfnamefont {M.~G.}\ \bibnamefont {Paris}},\ }\bibfield  {title} {\bibinfo {title} {Quantum communication with photon-number entangled states and realistic photodetection},\ }\href {https://doi.org/https://doi.org/10.1016/j.physleta.2010.01.016} {\bibfield  {journal} {\bibinfo  {journal} {Physics Letters A}\ }\textbf {\bibinfo {volume} {374}},\ \bibinfo {pages} {1342} (\bibinfo {year} {2010})}\BibitemShut {NoStop}%
\bibitem [{\citenamefont {Gumberidze}\ \emph {et~al.}(2016)\citenamefont {Gumberidze}, \citenamefont {Semenov}, \citenamefont {Vasylyev},\ and\ \citenamefont {Vogel}}]{Gumberidze2016}%
  \BibitemOpen
  \bibfield  {author} {\bibinfo {author} {\bibfnamefont {M.~O.}\ \bibnamefont {Gumberidze}}, \bibinfo {author} {\bibfnamefont {A.~A.}\ \bibnamefont {Semenov}}, \bibinfo {author} {\bibfnamefont {D.}~\bibnamefont {Vasylyev}},\ and\ \bibinfo {author} {\bibfnamefont {W.}~\bibnamefont {Vogel}},\ }\bibfield  {title} {\bibinfo {title} {Bell nonlocality in the turbulent atmosphere},\ }\href {https://doi.org/10.1103/PhysRevA.94.053801} {\bibfield  {journal} {\bibinfo  {journal} {Phys. Rev. A}\ }\textbf {\bibinfo {volume} {94}},\ \bibinfo {pages} {053801} (\bibinfo {year} {2016})}\BibitemShut {NoStop}%
\end{thebibliography}%
\nocite{*}
\end{document}